%% file: main.tex
\documentclass[sigplan]{acmart}
\renewcommand\footnotetextcopyrightpermission[1]{}
\AtBeginDocument{%
  \providecommand\BibTeX{{%
    \normalfont B\kern-0.5em{\scshape i\kern-0.25em b}\kern-0.8em\TeX}}}

\setcopyright{acmcopyright}
\copyrightyear{2018}
\acmYear{2018}
\acmDOI{XXXXXXX.XXXXXXX}

\acmConference[]{}{}{}
\acmPrice{15.00}
\acmISBN{978-1-4503-XXXX-X/18/06}

\usepackage{url}
\usepackage{breakurl}
\def\UrlBreaks{\do\/\do-\/\do_}
\makeatletter
\def\UrlAlphabet{%
      \do\a\do\b\do\c\do\d\do\e\do\f\do\g\do\h\do\i\do\j%
      \do\k\do\l\do\m\do\n\do\o\do\p\do\q\do\r\do\s\do\t%
      \do\u\do\v\do\w\do\x\do\y\do\z\do\A\do\B\do\C\do\D%
      \do\E\do\F\do\G\do\H\do\I\do\J\do\K\do\L\do\M\do\N%
      \do\O\do\P\do\Q\do\R\do\S\do\T\do\U\do\V\do\W\do\X%
      \do\Y\do\Z}
\def\UrlDigits{\do\1\do\2\do\3\do\4\do\5\do\6\do\7\do\8\do\9\do\0}
\g@addto@macro{\UrlBreaks}{\UrlOrds}
\g@addto@macro{\UrlBreaks}{\UrlAlphabet}
\g@addto@macro{\UrlBreaks}{\UrlDigits}
\makeatother

\usepackage{soul, color, colortbl}
\usepackage{amsmath}
\usepackage[ruled,vlined,linesnumbered]{algorithm2e}
\usepackage{comment}
\usepackage{graphicx}
\usepackage{svg}
\usepackage{caption}
\usepackage{subcaption}
\usepackage{threeparttable}
\usepackage{multirow}
\usepackage{booktabs}
\usepackage{verbatim}
\usepackage{epstopdf}
\usepackage{rotating}
\usepackage{listings}
\usepackage{listing}
\usepackage{paralist}
\usepackage{tabularx}
\usepackage{tabu}
\usepackage{arydshln}

\usepackage{amssymb}
\usepackage[shortlabels]{enumitem}
\usepackage{algpseudocode}
\usepackage{balance}
\usepackage{endnotes}
\usepackage{xspace}
\usepackage{amsfonts}
\usepackage{pifont}
\usepackage{wasysym}
\usepackage{tikz}
\usepackage{xcolor}
\usepackage{float}
\usepackage{multirow}

\usepackage{hyperref}
\definecolor{darkblue}{rgb}{0.0, 0.0, 0.55}
\definecolor{darkcandyapplered}{rgb}{0.64, 0.0, 0.0}
\hypersetup{
    colorlinks=true,
    linkcolor=darkblue,
    citecolor=darkcandyapplered,
    filecolor=magenta,      
    urlcolor=darkblue,
}
\newcommand{\ie}{\textit{i.e.}}

\definecolor{mauve}{rgb}{0.58,0,0.82}
\definecolor{mygray}{gray}{0.9}
\definecolor{dkgreen}{rgb}{0,0.6,0}

\newcommand{\Code}[1]{\lstinline{#1}}

\newcommand{\sys}{{\tt O2C}\xspace}

\begin{document}

\title{When eBPF Meets Machine Learning: On-the-fly OS Kernel Compartmentalization}

\author{Zicheng Wang}
\affiliation{ \institution{Nanjing University}  \institution{University of Colorado Boulder}
\country{}}

\author{Tiejin Chen}
\affiliation{\institution{Arizona State University}
\country{}}

\author{Qinrun Dai}
\affiliation{\institution{University of Colorado Boulder}
\country{}}

\author{Yueqi Chen}
\affiliation{\institution{University of Colorado Boulder}
\country{}}

\author{Hua Wei}
\affiliation{\institution{Arizona State University}
\country{}}

\author{Qingkai Zeng}
\affiliation{\institution{Nanjing University}
\country{}}

\begin{abstract}
Compartmentalization effectively prevents initial corruption from turning into a successful attack. 
This paper presents \sys, a pioneering system designed to enforce OS kernel compartmentalization on the fly.
It not only provides immediate remediation for sudden threats but also maintains consistent system availability through the enforcement process.

\sys is empowered by the newest advancements of the eBPF ecosystem which allows to instrument eBPF programs that perform enforcement actions into the kernel at runtime.
\sys takes the lead in embedding a machine learning model into eBPF programs, addressing unique challenges in on-the-fly compartmentalization.
Our comprehensive evaluation shows that \sys effectively confines damage within the compartment.
Further, we validate that decision tree is optimally suited for \sys owing to its advantages in processing tabular data, its explainable nature, and its compliance with the eBPF ecosystem. 
Last but not least, \sys is lightweight, showing negligible overhead and excellent sacalability system-wide.
\end{abstract}

\maketitle

\input{sections/introduction}

\input{sections/background}
\input{sections/motivate}
\input{sections/overview}

\input{sections/technique}
\input{sections/implementation}

\input{sections/evaluation}

\input{sections/discussion}
\input{sections/conclusion}

\newpage
{
\bibliographystyle{ACM-Reference-Format}
\bibliography{reference.bib}{}
}


\appendix

\input{sections/appendix}

\end{document}

%% file: sections/introduction.tex
\section{Introduction}
\label{sec:intro}
A successful attack is a multi-step process.
From an initial corruption, attackers need to leverage a sequence of exploitation techniques to obtain exploitable primitives, escalate privileges of the compromised system, until achieve the ultimate goals such as unauthorized access to sensitive information. 
Compartmentalization is an effective defense to prevent initial corruption from turning into a successful attack. 
It constrains the damage within the vulnerable component so that the rest of the system continues to operate well. 

A variety of compartmentalization solutions have been proposed for Operating System kernels.
Notable examples include methods that harness hardware features such as NOOKS~\cite{nooks03} and HAKC~\cite{mckeehakc22}, and those utilize hypervisors, including HUKO~\cite{xiong2011huko}, LXD~\cite{narayanan2019lxds}, LVD~\cite{narayananlvds20}, and KSplit~\cite{huangksplit22}.
Additionally, there are software-based approaches like SFI~\cite{wahbe1993sfi}, XFI~\cite{erlingsson2006xfi}, BGI~\cite{castro2009bgi} and LXFI~\cite{maolxfi11}.

Though these solutions can achieve the compartmentalization goal technique-wise, when applied to the real world, they are hindered from producing the best results.
On the one hand, security incidents often happen all of sudden~\cite{pwn2ownincident}, and there is no prior indication of which kernel component may be vulnerable. 
As a result, compartmentalization can be enforced only after the disclosure of security threats.
On the other hand, existing solutions, no matter whether hardware-based, hypervisor-based, or software-based, are all offline compartmentalization techniques. 
They require pre-reserving memory, pre-configuring hardware or hypervisor, or instrumenting kernel during compilation time. 
As a consequence, to enforce compartmentalization, we have no choice but to disrupt running computation services to recompile and reboot the system.

To maximize the protection effect of kernel compartmentalization, it is invaluable to have on-the-fly solutions that not only provide immediate remediation for sudden threats but also maintain consistent system availability.
However, achieving on-the-fly compartmentalization presents three unique challenges.
First, on-the-fly scenarios lack pre-arranged utilities: both hardware and hypervisor features require configuration before system bootup, and instrumentation must be done during compilation.
Second, the assets of the compartment and the rest of the kernel, such as stack and heap, are intertwined. It is difficult to clearly distinguish between them.
Third, at the beginning of compartmentalization, the compartment retains objects not tracked as they are allocated before time 0.
Maintaining data integrity under such uncertainties is challenging.

\begin{figure}
    \centering
    \includegraphics[width=.95\columnwidth]{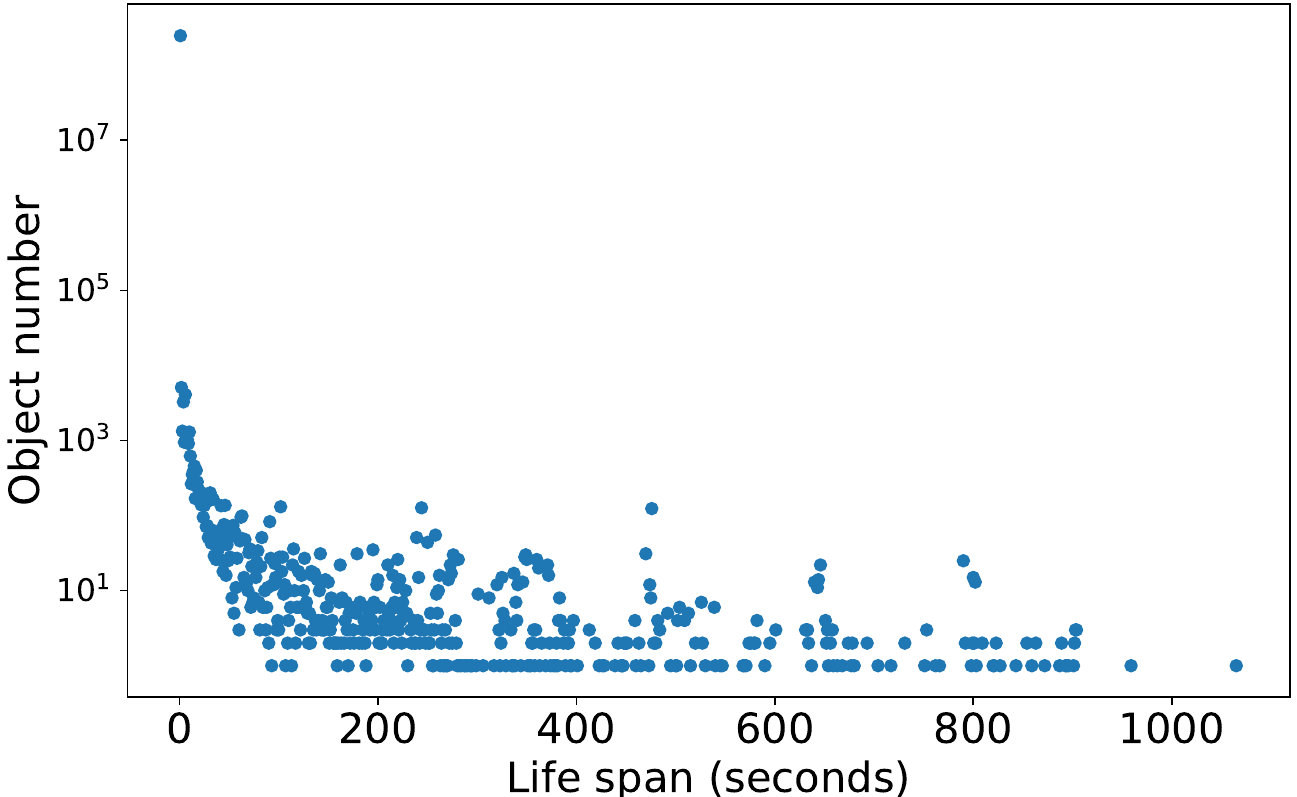}
    \vspace{-0.1in}
    \caption{Distribution of object, profiled in 20 minutes.}
    \label{fig:lifetime}
\end{figure}

Our profiling of kernel objects show that this transition phase is long enough for launching attacks: more than 3340 kernel objects have a lifespan longer than one minute (Figure~\ref{fig:lifetime}).   
Facing these challenges, on-the-fly kernel compartmentalization has been a mission impossible despite its various benefits.

In this work, we make a pioneering attempt by designing and implementing \sys. It addresses the three unique challenges through the newest advancements in the eBPF ecosystem and the wide application of machine learning. 
For the first challenge, \sys enforces compartmentalization using eBPF programs which can be instrumented into the kernel at runtime.
To tackle the second challenge,
\sys uses BPF maps and helper functions to create and manage private stack and heap. 
On the basis of this, \sys refines compartmentalization actions to achieve software fault isolation (SFI).
Addressing the third challenge, \sys innovatively embeds a machine learning model into the kernel space via eBPF programs, keeping data integrity throughout the transition phase.

We comprehensively evaluated \sys using real-world cases. 
The experiment results show that \sys brings in substantial security by preventing the damage within the compartment from jeopardizing the entire system.
We compared the performance of multiple machine learning models and validated that decision tree is the most suitable model in \sys, owing to its advantages in processing tabular data, its explainable nature, and its compliance with the constraints of eBPF ecosystem.
The performance measurement shows that system-wide, \sys has unnoticeable overhead and shows excellent scalability.
Specific to the impact on the compartment, \sys generally matches HAKC - the state-of-the-art hardware assisted offline compartmentalization solution, while each shows unique advantages under particular conditions.

To our knowledge, \sys is the first work on on-the-fly compartmentalization. It makes the following contributions: 

\begin{itemize}
    \setlength\itemsep{0.3em}
    \item Design and open-sourced implementation of \sys - the first solution to on-the-fly compartmentalization.
    \item Exploration of embedding machine learning models into the kernel space by utilizing the newest advancements in the eBPF ecosystem.
    \item A comprehensive evaluation of \sys which validates the effectiveness and efficiency of the design.
\end{itemize}

%% file: sections/background.tex
\section{Background and Related Works}

In this section, we will describe prior works on kernel compartmentalization and the eBPF ecosystem on top of which we design \sys.

\subsection{(Kernel) Compartmentalization}
For monolithic kernels like Linux, every kernel component, including loadable kernel modules and third-party device drivers, shares the same address space. 
Therefore, one single vulnerability, once exploited by attackers, can impede the entire kernel.
To practice the \textbf{Principle of Least Privilege}~\cite{uscope21,mckeehakc22}, research efforts have been undertaken to compartmentalize the untrusted kernel component, limiting the resources that it can access.
As such, we can constraint damages caused by vulnerability exploitation, and avoid affecting services provided by other kernel components.
The granularity of a compartment can range from entire device drivers from third-party vendors to kernel assets as fine as individual files and functions. 
One representative compartmentalization design consists of an access control policy and an enforcement mechanism. 

\vspace{0.05in}
\noindent \textbf{Policy.} 
The policy specifies which subject has what privilege to perform some set of operations on the data and resources accessible within a compartment. 
Typically, the kernel other than the compartment has the highest privilege, allowed to read and write to any memory regions and access any resources. 
In contrast, the compartment is de-privileged, constrained to using only the minimal set of memory and resources necessary to maintain its functionality.
It is prevented from compromising any other parts of the kernel.

\vspace{0.05in}
\noindent \textbf{Mechanism.} 
When the policy is ready, the compartmentalization relies on a mechanism to enforce it. 
The mechanism can be hardware-based~\cite{boyd2010sud,sun2013side} or hypervisor-based~\cite{xiong2013silver,srivastava2011gateway}. 
Landmarks include NOOKS~\cite{nooks03} and SKEE~\cite{azab2016skee} which leverage multiple kernel page tables, SCALPEL~\cite{dehonscalpel21} targeting the PIPE architecture, HAKC~\cite{mckeehakc22} which use the Arm architecture's new PAC and MTE features to provide enforcement, and SecureCells~\cite{bhattacharyya2023securecells} which designs specific CPU features for compartmentalization.
In addition, HUKO~\cite{xiong2011huko} employ hypervisor's extended page tables to compartmentalize the untrusted module, LXD~\cite{narayanan2019lxds}, LVD~\cite{narayananlvds20}, and KSplit~\cite{huangksplit22} further use \Code{vmfunction} to reduce performance cost.
These works require pre-reserving memory and pre-configuring the hardware or hypervisor before the kernel bootup.
Therefore, they cannot be used for on-the-fly compartmentalization.

The mechanism can also be software-based where memory access instructions are instrumented with checks during compilation time.
The checks will examine whether the accessd memory is within regions designated for the compartment.
Works in this direction are pioneered by SFI~\cite{wahbe1993sfi} and XFI~\cite{erlingsson2006xfi}, followed by BGI~\cite{castro2009bgi} which uses shadow memory to manage access control policies, LXFI~\cite{maolxfi11} which introduces a domain-specific language to maintain the integrity of interfaces between compartments and kernel.
Since memory pre-reserving is also needed in these works, we cannot directly apply them to the on-the-fly scenario either.

\vspace{0.05in}
\noindent \textbf{Other Compartmentalization.} 
In addition to compartmentalizing untrusted kernel, there are a series of works on compartmentalizing embedded system firmware~\cite{epoxy17,kimminion2018,aces18,khanm2mon21,dbox22,opec22,khanec23,khancrtc23} and libOS~\cite{hugo2022flexos,olivier2020unikernel}.
For instance, ACES~\cite{aces18} enforces developer-specified policies with MPU hardware feature for embedded system compartmentalization. FlexOS~\cite{hugo2022flexos} specializes the isolation strategy of a libOS at compilation/deployment time, and enforces strategies with multiple hardware and software protection mechanisms.

While \sys is designed to compartmentalize complex monolithic OS kernels, its on-the-fly capabilities can also be used in other compartmentalization scenarios.

\subsection{The eBPF Ecosystem}
\label{subsec:bk:ebpf}

The extended Berkeley Packet Filter (eBPF) is an in-kernel virtual machine that allows \textbf{privileged} users to run programs in the kernel space.
The eBPF programs are written in C language and compiled to bytecode using the LLVM eBPF backend. 
The compiled eBPF program is then instrumented to any kernel instructions as a dynamic extension. 
By extracting and updating execution context either before or after the instrumented instruction, the program monitor and modify kernel behaviors.

\vspace{0.05in}
\noindent \textbf{Pros.} 
Compared with other kernel extension mechanisms, particularly Kernel Probes~\cite{lwnTracingAttach}, the eBPF ecosystem has a range of pros.
First, the \textbf{safety} of eBPF programs is guaranteed by a static verifier which establishes three properties: (1) memory safety, ensuring that the program only accesses pre-defined memory locations, (2) information flow security, ensuring that no secrete kernel state is exposed, and (3) all execution must terminate.
Second, the eBPF programs show excellent \textbf{expressiveness} thanks to eBPF helper functions and BPF maps.
The eBPF helper functions allow the installed eBPF programs to interact with other kernel subsystems.
The BPF maps can store arbitrary data and can be used to share data across eBPF programs to with each other and privileged userspace processes.
Third, for the sake of \textbf{efficiency}, a variety of optimization techniques has been adopted in the eBPF ecosystem.
For instance, the eBPF bytecode is executed by a Just-In-Time (JIT) engine rather than a slow interpreter to achieve native-machine code level performance. 
Further, the instructions once attached, are overwritten to \Code{call} or \Code{jmp} instructions instead of the previously used \Code{int3} interrupt, which saves the time spent on context switching.


\vspace{0.05in}
\noindent \textbf{Cons.}
Despite so many pros, the current eBPF ecosystem has certain limitations. 
First, it lacks support for floating-point calculations which are fundamental operations in many applications, for example, machine learning techniques.
Second, the eBPF ecosystem restricts the maximum number of bytecode instructions in a single eBPF program.
Therefore, it cannot shoulder complex tasks.
Third, it disallows dynamic heap and has constraints on stack size, making it difficult to store data of large scale. 
Therefore, as we will discuss very soon, presently, the eBPF ecosystem is mainly used for traditional operating system services such as synchronization, storage, and networking.
There is limited research on embedding newest technique advancements, like machine learning, to eBPF programs for kernel tasks.

\vspace{0.05in}
\noindent \textbf{Applications.} 
In recent years, the eBPF ecosystem has undergone swift evolution across commodity OS kernels, including Linux, Windows, FreeBSD, and macOS~\cite{microsoftEBPFWindows,freebsdEBPFImplementation}.
There emerges a variety of eBPF applications, particularly in the field of networking, storage, and cloud computing in recent years~\cite{zhong2022xrp,he2022rapidpatch,park2022application,kaffes2021syrup,Rivittiehdl23,MarcohXDP20,MarcohXDP22,Yangelectrode23,Yanglambdaio23,Qispright23}.

For example, Syrup~\cite{kaffes2021syrup} employs eBPF to implement user-defined scheduling within the kernel. 
XRP~\cite{zhong2022xrp} and BMC~\cite{ghigoff2021bmc} utilize eBPF to enhance the performance of in-memory key-value stores. 
Syncord~\cite{park2022application} modifies kernel locks using eBPF.
$\lambda$-IO~\cite{Yanglambdaio23} improves computational storage devices by managing both computational and storage resources through eBPF. 
SPRIGHT~\cite{Qispright23} streamlines serverless computing by reducing protocol processing and serialization-deserialization overheads with eBPF. 
hXDP~\cite{MarcohXDP20,MarcohXDP22} and eHDL~\cite{Rivittiehdl23} synthesize eBPF programs to enable hardware functions in smart NICs.
In security area, PET~\cite{pet} instruments eBPF programs to error sites, preventing kernel vulnerabilities from being triggered.
RapidPatch~\cite{he2022rapidpatch} enables hot-patching of embedded devices using eBPF.

However, research into utilizing the eBPF ecosystem to program machine learning models for kernel tasks remains limited~\cite{bachl2022flowbased}, owing to the aforementioned cons of the eBPF ecosystem. 
One focus of this work is to explore this possibility in the scenario of on-the-fly compartmentalization.

%% file: sections/motivate.tex
\section{Motivating Example \& Challenges}
In this section, we will present a real-world security incident to illustrate why on-the-fly compartmentalization is needed, followed by the challenges of achieving it.

\subsection{A Security Incident}
In May 2022, Ubuntu was disclosed to be exploited through a use-after-free vulnerability~\cite{pwn2ownincident}.
The initial disclosure only pinpointed the vulnerability to the Linux kernel's traffic control networking subsystem, with no additional details provided.
Regardless of the limited information, immediate security actions were imperative for damage control.

Since a patch has yet to be released, applying one is infeasible.
Given the situation, compartmentalizing the traffic control networking subsystem becomes the most practical solution.
In fact, compartmentalization isn't just a temporary remedy for this specific incident.
It offers protections when, in the near future, the same subsystem was found to contain other 0-day vulnerabilities~\cite{route4blackhat-zhenpeng} and the corresponding patches prove to be incorrect~\cite{route4change-syz}.
However, to enforce the compartmentalization, existing techniques require to either reload the vulnerable module or power-off the machine, and then recompile and reboot the kernel, which inevitably disrupts system services.


\subsection{Challenges in On-the-fly Compartmentalization}
\label{lab:compartmentalization}

On-the-fly compartmentalization can address the problem by providing immediate remediation for sudden threats and in the meanwhile keeping services available. 
While its benefits are clear, it is challenging to achieve.

To thwart attacks stemming from the vulnerable kernel component, compartmentalization adheres to the principle of least privilege, ensuring the following properties:
\ding{182} \textbf{control flow integrity} -  the execution within the compartment cannot be hijacked to unauthorized target code.
To guarantee this property, compartmentalization needs to validate whether the destination of indirect calls and jumps and the stack's return address are legitimate.  
\ding{183} \textbf{data integrity} - the compartment is only allowed to access data owned by itself or co-owned with the kernel.
To upload this property, compartmentalization must scrutinize every memory access within the compartment, ensuring that the object accessed is in the right region and of the appropriate type.
\ding{184} \textbf{Authorized arguments and return value} - the compartment is forbidden from passing malicious arguments or returning malicious value to the remaining kernel because it can result in confused deputy attacks~\cite{lefeuvre2022civ} and Iago attacks~\cite{Checkoway2013iago}.

While mature solutions exist for achieving these goals in offline compartmentalization, when applying compartmentalization on the fly, three unique challenges emerge:
\begin{itemize}
    \setlength\itemsep{0.3em}
    \item \textbf{Mechanism Challenge (C1).} 
            On-the-fly scenarios lack pre-arranged utilities to enforce compartmentalization: both hardware and hypervisor features require configuration before system bootup, and instrumentation must be done during compilation. 
    \item \textbf{Expression Challenge (C2).} 
            Offline compartmentalization can pre-allocate private stack and heap to the target component.
            This establishes clear boundaries between the compartment and the rest of the kernel, making policy expression straightforward.
            However, in on-the-fly scenarios, the assets of the compartment and those of the rest of the kernel are intertwined, making distinction complex.
    \item \textbf{Transition Challenge (C3).}
            In offline scenarios, every data object is on record since system bootup. However, at the beginning of on-the-fly compartmentalization, the target component retains data objects allocated before time 0.
            These objects are not tracked and their types are unknown. Maintaining data integrity under such uncertainties poses a significant challenge.
\end{itemize}

%% file: sections/overview.tex
\section{Design Overview}

At a high level, \sys tackles the \textbf{mechanism challenge (C1)} by harnessing the eBPF ecosystem as it allows instrumenting eBPF programs into the kernel at runtime. 
To address the \textbf{expression challenge (C2)}, \sys leverages BPF maps and the helper functions to create and manage private stack and heap.
On the basis of this, \sys further refines compartmentalization actions to achieve software fault isolation (SFI).
For the \textbf{transition challenge (C3)}, \sys incorporates a machine learning model within the kernel to keep data integrity during transitions. 
In the following, we will present the security model and the access control policy of \sys, as well as its three-phase workflow.

\input{tables/policy}
\subsection{Security Model and Policy}

We assume the kernel is trustworthy as a whole, except for the component designated to be compartmentalized. 
Attackers don't have a root privilege so they cannot directly deactivate \sys. 
For the same reason, attackers cannot trigger and exploit potential vulnerabilities in the eBPF subsystem, the execution of which requires root privilege.
We deem the eBPF programs loaded by \sys are safe thanks to the sound verifier which provides formal assurance.

The goal of attackers is to exploit vulnerabilities in the compartment to breach other kernel parts and, ultimately, obtain root privileges or steal sensitive data.
They are allowed to use up-to-date exploitation techniques, excluding hardware side channels and physical attacks, as these are outside the scope of kernel-level compartmentalization.
The users of \sys are system administrators who have the right privilege to activate it.
\sys strives to adhere to the principle of least privilege, safeguarding other kernel parts from potential threats arising from vulnerabilities in the compartment, while simultaneously, ensuring that system services maintain effective functionality.

Table~\ref{tab:rwx} presents the access control policy enforced by \sys.
In this policy, the compartment is de-privileged and is deprived of permission to write to kernel data, heap, and stack.
Any operations that could potentially interact with the kernel, such as function calls and memory access, must undergo rigorous checks of eBPF programs instrumented to the compartment by \sys.


\begin{figure*}[t]
    \centering
    \includegraphics[width=.85\linewidth]{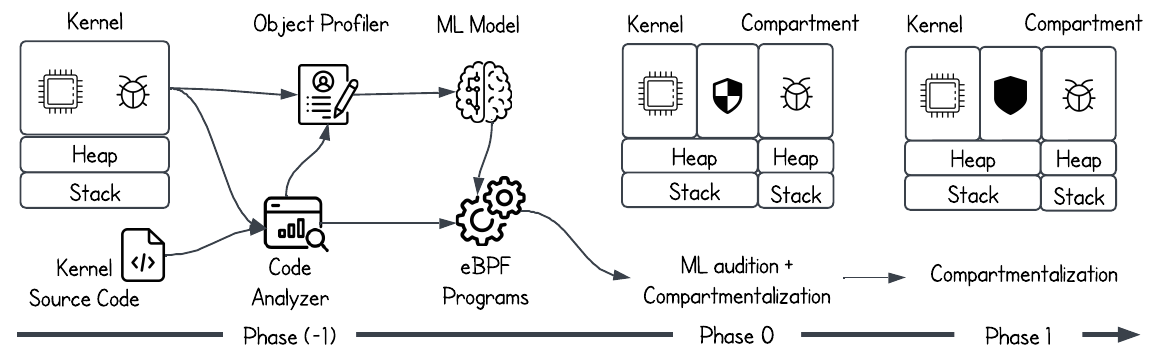}
    \caption{Three-phase workflow of \sys. In Phase (-1), \sys lays the groundwork. In Phase 0, compartmentalization is initialized yet require ML-based audition for untracked objects. Finally, in Phase 1, all objects are tracked, obviating the need for audition.}
    \label{fig:workflow}
\end{figure*}


\subsection{Three-phase Workflow}

\sys works across three phases: Phase (-1) for preparation, Phase 0 for transition, and Phase 1 finally establishes compartmentalization. Figure~\ref{fig:workflow} illustrates the process. 

In Phase (-1), the kernel runs as usual without the presence of threats, and \sys lays the groundwork for compartmentalization \footnote{The preparation can be conducted in an identical and assured safe machine.}. 
The Code Analyzer investigates the kernel source code and its runtime binary, identifying instructions - such as memory access - that require compartmentalization enforcement. 
It also facilitates the Object Profiler to label the content of kernel objects with the object type - the training data for machine learning (ML) model.
The model is able to infer the object type based on its content without tracing records. 
All enforcement actions are finally synthesized into a set of eBPF programs to be used in following phases.

Phase 0 transitions the vulnerable component from a non-compartmentalized state to a compartmentalized one.
Throughout this phase, \sys instruments eBPF programs to the instructions previously identified by the Code Analyzer in Phase (-1), creating private stack and heap, taking control over the subject transfer, and validating compartmentalization properties at corresponding sites.
In particular, to tackle the \textbf{transition challenge (C3)}, 
\sys performs audition by leveraging the ML model produced and embedded into eBPF programs in Phase (-1).
When detecting abnormalities, \sys logs essential information for postmortem analysis. 

Finally, in Phase 1, all objects within the compartment are tracked.
By this juncture, the ML-based audition becomes redundant and is thus deactivated.
Nevertheless, all other compartmentalization actions keep functioning.

%% file: tables/policy.tex
\definecolor{mygray}{gray}{0.9}
\begin{table}[t]
\resizebox{\columnwidth}{!}{%
\renewcommand\arraystretch{1.2}
\begin{tabular}{|l|ccl|cll|}
\hline
\textbf{}     & \multicolumn{3}{c|}{\textbf{Kernel}}                                                                          & \multicolumn{3}{c|}{\textbf{Compartment}}                                                                         \\ \cline{2-7} 
\textbf{}     & \multicolumn{1}{c|}{\textbf{read}} & \multicolumn{1}{c|}{\textbf{write}} & \multicolumn{1}{c|}{\textbf{exec}} & \multicolumn{1}{c|}{\textbf{read}} & \multicolumn{1}{c|}{\textbf{write}} & \multicolumn{1}{c|}{\textbf{exec}} \\ \hline
Kernel Code   & \multicolumn{1}{c|}{\checkmark}    & \multicolumn{1}{l|}{}               & \multicolumn{1}{c|}{\checkmark}    & \multicolumn{1}{c|}{\checkmark}    & \multicolumn{1}{l|}{}               &                                    \\ \hline
Kernel Data   & \multicolumn{1}{c|}{\checkmark}    & \multicolumn{1}{c|}{\checkmark}     &                                    & \multicolumn{1}{c|}{\checkmark}    & \multicolumn{1}{l|}{}               &                                    \\ \hline
Kernel Heap   & \multicolumn{1}{c|}{\checkmark}    & \multicolumn{1}{c|}{\checkmark}     &                                    & \multicolumn{1}{c|}{\checkmark}    & \multicolumn{1}{l|}{}               &                                    \\ \hline
Kernel Stack  & \multicolumn{1}{c|}{\checkmark}    & \multicolumn{1}{c|}{\checkmark}     &                                    & \multicolumn{1}{c|}{\checkmark}    & \multicolumn{1}{l|}{}               &                                    \\ \hline
\rowcolor{mygray} 
eBPF Programs & \multicolumn{1}{c|}{\checkmark}    & \multicolumn{1}{c|}{\checkmark}     & \multicolumn{1}{c|}{\checkmark}    & \multicolumn{1}{c|}{\checkmark}    & \multicolumn{1}{l|}{}               &                                    \\ \hline
Compartment Code  & \multicolumn{1}{c|}{\checkmark}    & \multicolumn{1}{c|}{\checkmark}     & \multicolumn{1}{c|}{\checkmark}    & \multicolumn{1}{c|}{\checkmark}    & \multicolumn{1}{l|}{}               & \multicolumn{1}{c|}{\checkmark}    \\ \hline
Compartment Data  & \multicolumn{1}{c|}{\checkmark}    & \multicolumn{1}{c|}{\checkmark}     &                                    & \multicolumn{1}{c|}{\checkmark}    & \multicolumn{1}{c|}{\checkmark}     &                                    \\ \hline
Compartment Heap  & \multicolumn{1}{c|}{\checkmark}    & \multicolumn{1}{c|}{\checkmark}     &                                    & \multicolumn{1}{c|}{\checkmark}    & \multicolumn{1}{c|}{\checkmark}     &                                    \\ \hline
Compartment Stack & \multicolumn{1}{c|}{\checkmark}    & \multicolumn{1}{c|}{\checkmark}     &                                    & \multicolumn{1}{c|}{\checkmark}    & \multicolumn{1}{c|}{\checkmark}     &                                    \\ \hline
\end{tabular}%
}
\vspace{0.05in}
\caption{Access control policy enforced by \sys, with two subjects: Kernel (including eBPF programs) and Compartment. The \checkmark~indicates that the subject has permission to read, write, or execute the corresponding object.}
\label{tab:rwx}
\vspace{-0.2in}
\end{table}

%% file: sections/technique.tex
\section{Code Analyzer and eBPF Programs}
\label{sec:analyzer}

\begin{figure}
    \centering
    \includegraphics[width=\linewidth]{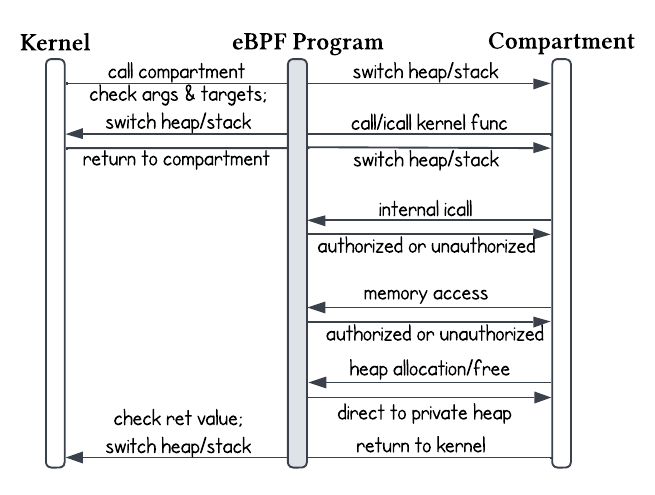}
    \vspace{-0.3in}
    \caption{The interaction protocol among the kernel, the eBPF programs, and the component.}
    \label{fig:switch}
\end{figure}

Now, we describe more details of the Code Analyzer and enforcement actions taken by eBPF programs to achieve SFI.
Figure~\ref{fig:switch} summarizes the interaction protocol among the kernel, the eBPF programs, and the compartment. 




\subsection{Identifying Instructions for Enforcement}
\label{sec:Identifying}
\sys instruments eBPF programs to the runtime kernel, scrutinizing three categories of instructions.


\vspace{0.05in}
\noindent \textbf{Indirect Transfer.} 
The first category is related to indirect execution transfer, including indirect calls, indirect jumps, and function returns.
In Phases 0 and 1, \sys meticulously examines the target addresses of these instructions to make sure that they conform with the component's legitimate control flow graph.
To this, when identifying these instructions in Phase (-1), the Code Analyzer not only records their own address in the format of \Code{func+offset} but also marks where the target address is stored. 
For example, for instructions \Code{call *\%rax}, the Code Analyzer logs that the target address can be retrieved from \Code{rax} register, and for \Code{jmp *-0x7dabaac0(,\%rax,8)}, the target address resides in the \Code{(0x7dabaac0+\%rax*8)} memory cell.
In particular, for \Code{ret} instructions, the target address is implicitly placed in the \Code{\%rbp+0x10} for x86 architecture, and in the \Code{\%r14} register for ARM.


\vspace{0.05in}
\noindent \textbf{Memory Access.} 
The second category of instructions pertains to memory access.
These instructions include move, store, load, as well as comparison and arithmetic operations.
Either the source or destination operands of them may involve memory.
Furthermore, the set includes \Code{push/pop} instructions that manipulate the kernel stack.
Similar to the first category, the Code Analyzer records both the instruction address and the accessed memory address. 
For example, the accessed address is \Code{\%rip + 0x29c23e4} in the instruction \Code{mov 0x29c23e4(\%rip),\%eax}, which denotes a global variable.
When the stack is accessed, the address is calculated from \Code{rbp/rsp} registers, as they represent the bottom and top of the stack, respectively.
In Phases 0 and 1, \sys evaluates whether the access adheres to compartmentalization properties.
Briefly, it checks whether the addresses fall within private stack and heap, except for those co-owned with the kernel.
To support private heap, the Code Analyzer also identifies \Code{call} to heap allocation.
\sys intercepts the allocation request and obtains memory from the private heap beginning from Phase 0.

\vspace{0.05in}
\noindent \textbf{Subject Switching.} 
The last category relates to subject switching, which happens when the compartment calls to functions in the rest of the kernel, and vice versa.
On one hand, since the compartment is untrusted, \sys needs to prevent it from passing malicious arguments or returning malicious values to external kernel functions, which could otherwise lead to harm.
On the other hand, the compartment operates on its own private heap and stack, ensuring that any internal corruption doesn't spread outwardly.
Therefore, as the subject switches, \sys should also switch stack and heap. 
To this end, the  Code Analyzer identifies entries of the functions that can be called from outside, as well as instructions that call out\footnote{The instruction next to \Code{call} is also identified so that \sys can switch heap and stack back to private ones after the external function returns.}.
Specifically, \sys creates a private stack when compartment functions are called for the first time.


\subsection{Enforcement Actions via eBPF programs}
\sys instruments eBPF programs to identified instructions, enforcing the access control policy in Table~\ref{tab:rwx}.


\vspace{0.05in}
\noindent \textbf{Control-flow Integrity.}
\sys guarantees the control-flow integrity by instrumenting a set of eBPF programs to instructions in the indirect transfer category.
More specifically, the Code Analyzer employs state-of-the-art static analysis techniques recently developed in~\cite{lu2019mlta,Sungbae2022kcfi} to obtain a sound and sufficiently accurate control flow graph. 
For each indirect call, indirect jump, and return instruction, this control flow graph provides a set of legal targets. 
\sys uses a BPF map to store this information, setting the instruction address as the key, and an array of targets as its value.
This map is shared among all eBPF programs dedicated to control-flow integrity.
By retrieving the runtime target, for example, from \Code{rax} register,  the eBPF program compares it with those stored in the map, determining whether the execution transfer breaches integrity or not.

\vspace{0.05in}
\noindent \textbf{Private Stack and Heap.}
\sys creates a private stack and a private heap for the compartment, which facilitates data integrity enforcement.
By instrumenting eBPF programs into instructions pertaining to memory access and subject switching, \sys restricts memory access to these private regions.
These private regions are referenced through shared BPF maps and managed by eBPF programs distributively.

In Phase 0, when execution enters the vulnerable component for the first time, the eBPF program instrumented to the function entry is executed. 
This program creates a private stack by obtaining memory from the buddy system which manages the entire computer's physical memory. 
The address of the stack is recorded in the BPF map so that it can be retrieved and reused in the future.

Similarly, to manage the private heap, \sys instruments eBPF programs to instructions that \Code{call} to allocation and free operations.
These programs conceptually form an internal allocator specific to the compartment.
At allocation sites, depending on what memory type is needed, the eBPF program takes different actions.
If multiple physically continuous pages are needed, the eBPF programs divert the allocation request to the buddy system, just like the creation of the private stack.
If an object smaller than one page is allocated, the eBPF program turns to the SLAB/SLUB allocator.
Inside the SLAB/SLUB allocator, memory is organized in the unit of slab caches. 
One slab cache is divided into uniformly sized slots so that objects in the same cache are either of the same type or of similar size.
When an object of a specific type is allocated for the first time, the eBPF program creates a private cache first, and from the private cache, the eBPF program uses a slot to hold the object. 
The address of the private cache is recorded in the shared BPF map so that the following allocations can come from it.
For objects that are co-owned by both the compartment and the kernel, without violating the access control policy, the eBPF program deems the cache holding the object as private.
If large blocks of virtually continuous memory are needed, the eBPF programs divert the request to the vmalloc allocator.
At free sites, \sys does not recycle the memory back to allocators but instead retains them for future reuse.
As such, the dynamic memory remains private to the compartment.


\vspace{0.05in}
\noindent \textbf{Data Integrity.}
With private stack and heap, \sys is able to perform data integrity checks at memory access instructions.
If the Code Analyzer in Phase (-1) determines that the accessed address should belong to a global or stack region, any deviation at runtime indicates malicious activity.
If the accessed address falls within a heap region, the Coder Analyzer further harnesses the object-alias static analysis developed in ~\cite{lu2022goshawk, lu2023typm} to identify the object type \footnote{The Code Analyzer annotates ordinary arrays with a unique ID as the type}.
In Phases 0 and 1, \sys has different designs to examine if the type of accessed objects at runtime matches the analysis results.

Since the heap objects in Phase 1 (and also some objects in Phase 0) are tracked in the shared BPF map, \sys directly consults it to validate the object type.
To break down, for objects allocated by the buddy system, the shared BPF map records its allocation site which acts as a type identifier.
For objects in slab caches, \sys looks up the shared BPF map to get a collection of cache structures where the object can legally reside.
From cache structures in this collection, \sys extracts the type information for comparison.
For those allocated by the vmalloc allocator, the shared BPF maps store the address of corresponding \Code{struct vm_struct} which can also tell the type.
However, in Phase 0, some objects have lifecycles starting in Phase (-1) and thus are untracked.
For these, \sys resorts to machine learning techniques, which will be detailed in Section~\ref{sec:ml}.

\vspace{0.05in}
\noindent \textbf{Authorized Arguments and Return Values.}
To forbidden confused deputy attacks~\cite{lefeuvre2022civ} and Iago attacks~\cite{Checkoway2013iago}, \sys supports runtime checks on both arguments and return values, where data flows from the compartment to the kernel.
Due to the complexity and diversity of the interface, the legal value range for arguments and return values can differ substantially.
For example, the expected return value for a file open operation should be either 0 to indicate success or \Code{-ERROR CODE} for failure, while that for file read should be the number of bytes read.
Prior compartmentalization works like LXFI~\cite{maolxfi11}, LVD~\cite{narayananlvds20}, and KSplit~\cite{huangksplit22} rely on annotation with assistance from interval analysis~\cite{wang2012kint} to compute the legal value range. 
\sys builds upon these efforts, providing a more direct approach - expressing the checks in eBPF programs.

\subsection{Optimization for Efficiency}
\label{lab:optimization}




\sys employs a variety of optimizations to keep lightweight.
These optimizations lead to a nearly 54\% reduction in the number of instructions requiring probes, while also shortcut most heap checks to basic comparison operations.


\vspace{0.05in}
\noindent\textbf{Skip Deterministic Addresses.} 
Instructions that access global kernel variables often appear as \Code{mov offset(\%rip), \%rax}. 
Therefore, the accessed address is always fixed when the instruction is being executed.
This certainty reduces the need for \sys to instrument global variable access instructions, accounting for 0.74\% of the total instrumentations.

Similar optimization can also be applied to instructions accessing stack variables.
In the prologue, a function's stack frame is created through \Code{sub offset1, \%rsp}, and the value of \Code{\%rbp/\%rsp} remains unchanged throughout the function's lifecycle.
Hence, instructions like \Code{mov x, offset2(\%rbp/\%rsp)} which have a fixed offset from \Code{\%rbp/\%rsp} are for sure accessing the current function's stack, obviating the need for additional runtime checks.
However, in some corner cases, such as \Code{\%rsp + \%rax * offset3}, where the accessed address depends on the register value at runtime, \sys retains check to guarantee that the accessed part is within the private stack.
This optimization leads to a reduction of 7.6\% in the total instrumentations and 99.32\% in stack access instrumentation.

\vspace{0.05in}
\noindent\textbf{Eliminate Redundant Checks.}
Because of the principle of locality, sequentially accessed memory addresses tend to be proximate, sharing the same base address but differing in their offsets.
It frequently arises during array indexing and operations on data structure fields.
To leverage this, we consolidate multiple access checks into a single check in which the eBPF program examines the minimum and maximum offsets - representing the two extremities.

Another redundant check is also related to stack access.
Because return address overwriting results from stack corruption and stack accesses have already undergone stringent scrutiny, adding an additional integrity check for the return address becomes superfluous.
Therefore, we can safely eliminate this type of instrumentation by 6.42\%.

\vspace{0.05in}
\noindent\textbf{Eschew Low-risk Read.}
In prior compartmentalization works, it is widely accepted that memory overreads are of low risk~\cite{cvss,wahbe1993sfi,castro2009bgi,erlingsson2006xfi,maolxfi11}, while their occurrence is twice as frequent as memory writes.
Since there is no dissent on this matter in the literature, without compromising security, \sys also opts not to scrutinize memory reads, and reduces 38.66\% of instrumented instructions.


\vspace{0.05in}
\noindent\textbf{Shortcut Heap Checks.}
Experimental results from object equivalence analysis~\cite{periklis2008wit,bhatkar2008dsr,belleville2018hardware,xu2022regvault} and lossless dynamic analysis~\cite{roessler2021lossless,uscope21} reveal that most heap access only writes to one object type. 
In such cases, \sys directly records the address of the specific cache structure to the shared BPF map, bypassing the step of first retrieving the collection of cache structures and then comparing them one by one.
As such, the check process is accelerated.



\section{eBPF-powered Machine Learning}
\label{sec:ml}


In this section, we address the \textbf{transition challenge (C3)}.
That is, in Phase 0, some heap objects of the compartment have lifecycles starting in Phase (-1).
They are not tracked and their types are unknown, making it difficult to examine data integrity.
This challenge hinders prior works from achieving on-the-fly compartmentalization.

\sys tackles this problem by embedding machine learning models to eBPF programs.
The model is trained to learn invariants in various kernel object types~\cite{arati2008invariants}.
At runtime, given a pointer dereference site, the model extracts the content of the referred heap object and infers its type.
If the inferred type doesn't match the static analysis results from the Code Analyzer, \sys auditions this malicious activity for postmortem analysis which is conducted asynchronously for the sake of performance.

We choose to perform audition at free sites rather than access sites.
This choice is motivated by the observation that the object's content is either zeroed out or filled with trash value right after allocation, rendering it indistinguishable from other objects. 
As the kernel keeps updating the  object throughout its lifecycle, its content progressively accumulates values specific to its type, making it increasingly distinctive.
Thus, by the time the object is freed, its content is mature enough for inferring.



\subsection{Collect Training Data via Object Profiler}

The Object Profiler in \sys collects training data in Phase (-1) when threats are not present. 
It employs a set of eBPF program to extract the content of heap objects.

Technically, at every allocation site, an eBPF program is instrumented to record the address of the allocated object as well as the call trace.
Analyzing the call trace, the Code Analyzer obtains the type of the allocated object.
Then, both the address and the type is stored to a BPF map.
At every free site, another eBPF program is instrumented to dump the content of the freed object. 
By looking up the address of the freed object in the BPF map, the eBPF program identifies the type and label the content accordingly. 
The labeling also takes place at free sites for the same reason we just discussed.

To ensure enough data for training, the collection is not restricted to the component to be compartmentalized but spans globally across the kernel.




\subsection{Decision Tree Fits Best}

There are numerous machine learning models available, such as support vector machines and neural networks, for the task of inferring object types based on object content. 
Among the various options, we choose decision tree as our default model based on the following insights. 

First, in comparison to the state-of-the-art deep learning methods, the decision tree model has distinct advantages in handling tabular data~\cite{shwartz2022tabular}. 
Our evaluation (Section~\ref{subsec:eval:ml}) compared decision tree model to other models, validating its efficacy for our specific task.
Second, the decision tree model is explainable and has deterministic inference time~\cite{molnar2020interpretable}, rendering it sufficiently trustworthy to be used in the kernel whose reliability and prompt response are paramount.
Third, as discussed in Section~\ref{subsec:bk:ebpf}, the current eBPF ecosystem imposes several constraints, including lacking support for floating-point calculation and dynamic heap, as well as limitations on the maximum number of instructions and stack size.
Given these restrictions, the decision tree is more viable for eBPF-based implementation than complex tree ensemble machine learning models such as Random forest~\cite{breiman2001random} or XGboost~\cite{chen2016xgboost}.
In Section~\ref{sec:discussion}, we will discuss our future plan to extend the eBPF ecosystem to embrace more machine learning models.

When training the model, given that the size of some objects are up to two pages, we split the object content into a sequence of quad words to form the feature vector, instead of training over the entire content. 
While this approach inflates the input dimension, the decision tree is adept at high-dimension data.  
Further, to deal with the varying sizes of objects, we standardize their lengths by padding zero values.

To embed the model into eBPF programs, we leverage \textit{scikit-learn}~\cite{buitinck2013api} to convert the model to five arrays:  \textit{childrenLeft}, \textit{childrenRight}, \textit{feature}, \textit{threshold}, and \textit{value}. 
The childrenLeft and childrenRight arrays build the tree structure. 
The feature array specifies which quad word in the feature vector is to be used at a tree node.
The threshold array indicates the split value at a tree node. 
Recall that eBPF programs don't support floating-point calculation, the decision tree model shows unique advantage in bypassing this restriction.
Inspired by prior work~\cite{bachl2022flowbased}, we round down the float threshold to an integer without losing precise.
This works because all comparisons in the decision tree are $\leq$ operation and all quad words in our model's feature vector are integers.
Last, the value array stores the classification output, \ie, object type, at leaf nodes.
The five arrays are loaded into BPF maps which are shared among all instrumented eBPF programs.
These programs execute the tree traversal logic and determine if the inferred type matches with the results from the Code Analyzer.

In addition to overcoming the constraints on floating-point calculation, we also address the issue of limited BPF stack size and lacking support for dynamic heap. 
To use the decision tree model, the eBPF program needs to extract the content of heap objects.
However, the size of some objects is larger the stack size, and dynamic heap is not present in the current eBPF ecosystem either.
To solve this, we pre-reserve a huge pool in a designated BPF map and store the runtime content in the map. 
With a pointer referring to the object in the pool, we read in the feature vector and fed it to the decision tree for type inference.

%% file: sections/implementation.tex
\section{Implementation}
The prototype of \sys includes 324 lines of C code for eBPF program templates, 260 lines of C code for the new helper functions in the kernel, and 1371 lines of Python code for identifying instructions for enforcement, machine learning training, and framework integration.
In addition, 8283 lines of C++ code for static analysis. 

\sys is implemented over Linux for its open-source nature. 
It can be effortlessly migrated to other mainstream OS kernels due to eBPF's consistent design across platforms.
Once the paper is accepted, all code will be open-sourced in GPL License.

\vspace{0.05in}
\noindent \textbf{Identifying Instructions for Enforcement.}
We employed Ghidra~\cite{ghidra} to disassemble kernel image. As a full-fledged reverse engineering tool, Ghidra can offer a full spectrum of analysis results, including sections, function boundaries, cross-references, variable identification, and more.

Starting from the lowest address in the compartment, we traverse all functions and use the \Code{CodeUnits} iterator and each instruction's mnemonic provided by Ghidra to classify instructions.
We identify instructions with \Code{mov, xchg, stos, out, rep} mnemonics as potential memory access instructions. 
Further, according to x86 convention, the two operands cannot both be memory addresses.
Thus, we can distinguish between reading from and writing to memory based on the operand type.
For execution transfer, the difference between direct call and indirect call is that the \Code{operand-0}'s type of the former is either \Code{ADDRESS} or \Code{CODE}, and the remaining cases can be classified as indirect calls.

\vspace{0.05in}
\noindent \textbf{Enriching Training Data.}
To enrich training data for the machine learning model, \sys employs Syzkaller~\cite{syzkaller} - the fuzzing tool developed by Google to proactively execute kernel.
Besides, \sys uses object-driven fuzzing techniques proposed in  GREBE~\cite{grebe} to focus kernel execution on operating objects of interest. 
By varying the operation context, we can diversify object content and thus obtain high-quality training data.

\vspace{0.05in}
\noindent \textbf{Helper Functions and BPF Maps.}
The current eBPF ecosystem has no support for register set and memory allocation. 
To support private heap and stack, we add four new helper functions with sufficient runtime checks.
The first is \Code{bpf_set_regs} which can set register values, including stack registers \Code{rsp/rbp}. Using this helper function, \sys can switch stacks.
The other three helper functions are for private heap, they are \Code{bpf_create_slab_cache} which creates private slab caches for the compartment and \Code{bpf_cache_alloc/free} which allocates from and frees to private caches.
Except for the four helper functions essential to \sys's functionality, we added another set of helper functions to simply operations.
For example, \Code{bpf_get_slab} and \Code{bpf_get_vmstruct} can get the description of the slab and vmalloc directly, without the need to traverse slab pages or \Code{vm_struct} rb-trees.

\sys utilizes various types of BPF maps to meet specific needs. 
In addition to using \Code{BPF_MAP_TYPE_ARRAY} for storing arrays essential to decision tree models, \sys predominantly relies on hash tables, denoted as \Code{BBPF_MAP_TYPE_HASH} for SFI.
These hash tables are used for locating slab caches, vmalloc vm\_struct, and buddy objects 
Furthermore, in the Object Profiler, \sys employs \Code{BPF_MAP_TYPE_RINGBUF}, a user-kernel communication ring buffer, to collect the content of heap objects.

%% file: sections/evaluation.tex
\section{Evaluation}
\label{sec:evaluation}




In this section, we comprehensively evaluate \sys using real-world cases.
We perform a security analysis of \sys, followed by the comparative experiments of machine learning models.
Then, we measure \sys's overhead.


\subsection{Security Analysis}
We investigated how \sys mitigates attack vectors outlined in our security model.
Furthermore, to evaluate the effectiveness of \sys in real-world situations, we collected 84 vulnerabilities from the IPv6, net/sched, and netfilter modules - three modules that are notably vulnerable~\cite{syzkaller}.

\vspace{0.05in}
\noindent\textbf{Control Flow Integrity.}
Attackers exploiting vulnerabilities within a compartment can hijack control flow to perform code-reuse attack and finally compromise the entire system.
To counteract this, \sys ensures that all targets of indirect calls and jumps are legal.
Besides, \sys examines all writes to the stack in the compartment to prevent out-of-bound writing.
As such, any intention to tamper with the return value on stack is caught before it can cause any adverse impact.

\vspace{0.05in}
\noindent\textbf{Data Integrity.} 
In addition to control flow hijacking, attackers can launch data-only attack~\cite{davi2017pt} by tampering with kernel code, global data, local variables on stack, and security-sensitive objects on heap.
\sys thwarts this threat by instrumenting eBPF programs to all memory writes for clearance.
Specifically, the kernel code is safeguarded by W$\otimes$X, and the compartment is forbidden from modifying the CPU status and page table entries.
For global data, the compartment is only permitted to access what is allowed under legitimate circumstances, adhering to the principle of least privilege.
\sys creates a private stack to the compartment and the kernel stack remains hidden during its execution.
Lastly, the compartment's access to heap objects is under stringent regulation, and authorization is granted only for objects it owns or legally co-owns with the kernel.

\vspace{0.05in}
\noindent\textbf{Authorized Arguments and Return Value.}
System administrators, as users of \sys can employ interval analysis to annotate the legal value range for arguments and return value, and express the corresponding checks in eBPF programs. 
For example, in the Confused Deputy Attack misusing \Code{memcpy(dst, src, size)}, \sys guarantees that addresses of security-sensitive kernel objects are not maliciously passed as the destination argument.
When a compartment returns a value to the caller function in kernel, \sys makes sure that the returned value aligns with the function's intended behavior and can be well handled by the caller.

\input{tables/vulns}

\vspace{0.05in}
\noindent\textbf{Real-world Vulnerabilities.}
Our analysis of 400 vulnerabilities indicates that memory corruption~\cite{alldata}, especially those targeting kernel heap is the dominant threat to the Linux kernel.
We sampled eight representative vulnerabilities in Table~\ref{tab:vulns} to illustrate how \sys prevent vulnerabilities in the compartment from compromising the entire system.
In the table, CVE-2021-3715 and CVE-2022-2588 are heap use-after-free vulnerabilities, and CVE-2022-27666 is a heap overflow vulnerability. 
\sys mitigates these threats by restricting the compartment's access to kernel objects of right type.
CVE-2023-0394 and syz-76d0b8 access unauthorized kernel addresses and thus trigger a CPU general protection fault.
\sys counters this by validating the target addresses for memory access.
Integer overflow vulnerabilities, such as syz-490321, and stack overflow vulnerabilities like syz-e73923, have the potential to cause kernel stack overflows and tamper with function return addresses. 
\sys creates private stack for the compartment and uses stack access check to safeguard return addresses.
Lastly, syz-caed28 exemplifies invalid free, where a pointer referring to incorrect type is passed as free's argument. 
\sys prevents this by enforcing argument authorization on the \Code{kfree} function.
In summary, \sys delivers comprehensive defenses against a variety of real-world vulnerabilities.


\vspace{0.05in}
\noindent\textbf{ML Audition.}
Since the ML audition takes place at free sites rather than access sites, it is possible that the vulnerability may have been triggered and exploited in the time window.
This malicious activity still undergoes audition.
Furthermore, if attackers deliberately disable audition after a successful penetration, the disabling itself is identified as an indicator of attacks.
If attackers attempt to evade audition by keeping objects alive long enough through Phase 0, their efforts will be thwarted because finally in Phase 1, \sys can still detect the abnormality.

\input{tables/ml-compare-to-random}
\input{tables/ml-parameters}

\subsection{Machine Learning Model Evaluation}
\label{subsec:eval:ml}

\vspace{0.05in}
\noindent\textbf{Experiment Setup.} 
We evaluated the performance of machine learning models through cross-validation.
Considering the numerous object types, traditional cross validation process will miss certain types in model training.
Therefore, we employed 5-fold Stratified cross-validation to include all object types in training set.
Regarding metrics, we consider not only accuracy but also Macro F-1 score which measures the model's performance over object types that have limited training samples.

We trained model instances with two granularities. One outputs specific object types (Type-Granularity) and the other directly tells whether the type of accessed object belongs to the compartment (Compartment-Granularity).
Tables~\ref{tab:ml-data} and \ref{tab:ml-features} describe the training data set and its features and labels.

\input{tables/ml-dataset}

To illustrate why the decision tree is the most suitable model for our specific task,  we compared the decision tree model customized for eBPF programs, to an unrestricted random forest model and an unrestricted neural network.
In particular, the neural network~\cite{popov2019neural,katzir2020net} has three hidden layers with size of neurons of each hidden layer are 256,128 and 64. We used Adam as optimizer and train the neural networks 50 epochs.

Last, we adjusted the length of feature vector and the maximum depths to identify the optimal parameters for the decision tree.


\vspace{0.05in}
\noindent\textbf{Experiment Results.} 
Table~\ref{ml tab:main} shows that the decision tree's Type-Granularity instance achieving 80\% - 96\% accuracy, and its Compartment-Granularity instance exceeding 99\% accuracy, across all datasets. 
These results are comparable to those of the unrestricted random forest model - a more advanced version of decision tree.
Further, the Macro F1 scores for both instances are on par with the random forest too. 
This indicates that even after customizing the decision tree model for eBPF programs, it retains its efficacy over object types with limited data. 
In contrast, the Macro F1 score the neural network model is much lower because, as a multilayer perceptron, it cannot handle imbalance tabular data.
Considering that the random forest cannot be embedded due to the limitations of current eBFP ecosystem, the decision tree model stands out as the more suitable model for \sys.


Table~\ref{ml tab:depths} presents the impact of varying parameter values on the decision tree model's performance.
The results show that the length of the feature vector is not decisive and 256 bytes are good enough.
The critical parameter is the tree depth, particularly taking the Macro F1 score into account. 
A shallower depth causes the model to overlook object types with limited training samples.
Within the constraints imposed by the eBPF ecosystem, we determined that a tree depth of 14 produces optimal results.

Further, comparing the results of two instances in Tables~\ref{ml tab:main} and~\ref{ml tab:depths}, we can observe that the coarse-grained instance (Compartment-Granularity) has much higher accuracy.
However, in production scenario, this accuracy introduces the risk of allowing attackers to compromise data within the compartment. 
This risk arises because the pointer can refer to an error type as long as the error type is used in the compartment.
\sys provides both instances in its implementation, thereby enabling users to make tradeoff based on risk preference.

\subsection{Performance Overhead}
\label{lab:performance-overhead}
Through experiments, we aim to answer the following questions : 
\textbf{RQ1:} What is the overall performance overhead of \sys on the entire system?
\textbf{RQ2:} What performance overhead does \sys impose within the compartment itself?
When answering the questions, we varied scales of the compartment and evaluated performance under the presence and absence of ML audition.


\vspace{0.05in}
\noindent\textbf{Experiment Setup.}
We continue to compartmentalize IPv6, net/sched, and netfilter - the three most vulnerable subsystem in performance measurement. 
IPv6 functions Internet Protocol version 6, net/sched manages network packet scheduling and Quality of Service (QoS), and netfilter handles packet filtering and Network Address Translation (NAT). 

All measurement were conducted on a machine with CPU Intel core i9-13900HX, 64 GB memory, and 1TB disk, running Ubuntu-22.10 with kernel v6.1 - the newest Long-term Support (LTS) version when we did experiments. Further, we set the kernel to performance mode and locked the CPU frequency, ruling out noises caused by Intel Turbo Boost.

For \textbf{RQ1}, we used LMbench~\cite{mcvoy1996lmbench} to measure system calls and Phoronix Test Suites~\cite{phoronix} to run representative real-world applications, such as Perf, Apache/Nginx-IPv4, Git, SQLite, Redis, XZ compression, and kernel compilation.
We varied the scale compartmentalization for each compartment, from a single core file to the entire subsystem.
Adding the last scenarios where all three subsystems were compartmentalized, we had in total of seven distinct situations. 
For each of these situations, we further measured the performance with and without ML audition.

For \textbf{RQ2}, we selected IPv6 as the target subsystem so that we can draw comparison with HAKC - the state-of-the-art compartmentalization technique. 
Besides, IPv6 can be fully tested using Apache Bench in metrics of requests per second and transfer rate.
In more details, we booted a local IPv6 Web Server binding \Code{[::1]:8000} and compartmentalized IPv6 subsystem of the server. 
Through the IPv6 address, we accessed files of three different sizes - 100KB, 1MB, and 10MB - a total of 1000 times.

Note that we compared \sys with hardware-assisted HAKC rather than other SFI-based compartmentalization works such as XFI~\cite{erlingsson2006xfi}, BGI~\cite{castro2009bgi}, and LXFI~\cite{maolxfi11}.
This is because XFI and BGI are designed for Windows operating systems, rendering a direct comparison infeasible.
Regarding LXFI, its implementation is not open-sourced, and it compartmentalizes the e1000 Intel network driver, the corresponding NIC of which is outdated and is no longer on the market.

\input{tables/overhead}


\vspace{0.05in}
\noindent \textbf{Results for RQ1.}
Table~\ref{tab:overhead} presents sampled results for system-wide performance overhead of \sys, with vanilla kernel as the baseline. More complete results are moved to \cite{alldata} due to the space limits.

From the table, we can observe that the overall overhead of \sys ranges from -1.26\% to 3.81\% for LMbench, which is withing reasonable margin of fluctuation.
For Phoronix, the overhead is slightly higher, especially kernel compilation and Apache. 
This is because these applications either frequently create child processes or continuously process incoming network packets.
These operations result in numerous memory allocations, all of which need to be tracked by \sys, leading to performance loss.

Regarding the impact of ML audition, from the table, we can see that it is unnoticeable across all used benchmarks. 
This indicates a smooth transition from Phase 0 to Phase 1 as users won't experience discernible change inside kernel.
Further, as the scale of the compartment increases, beginning from a singe file \Code{ip6_output.c} with 1,470 lines of code (LOC), to the IPv6 subsystem with 78,213 LOC, and finally all three subsystems at 255,330 LOC, the system-wide performance doesn't show obvious degradation, attesting \sys's excellent scalability from the perspective of entire system.

\begin{figure}[t]
    \begin{subfigure}[b]{0.5\textwidth}
         \centering
         \includegraphics[width=\textwidth]{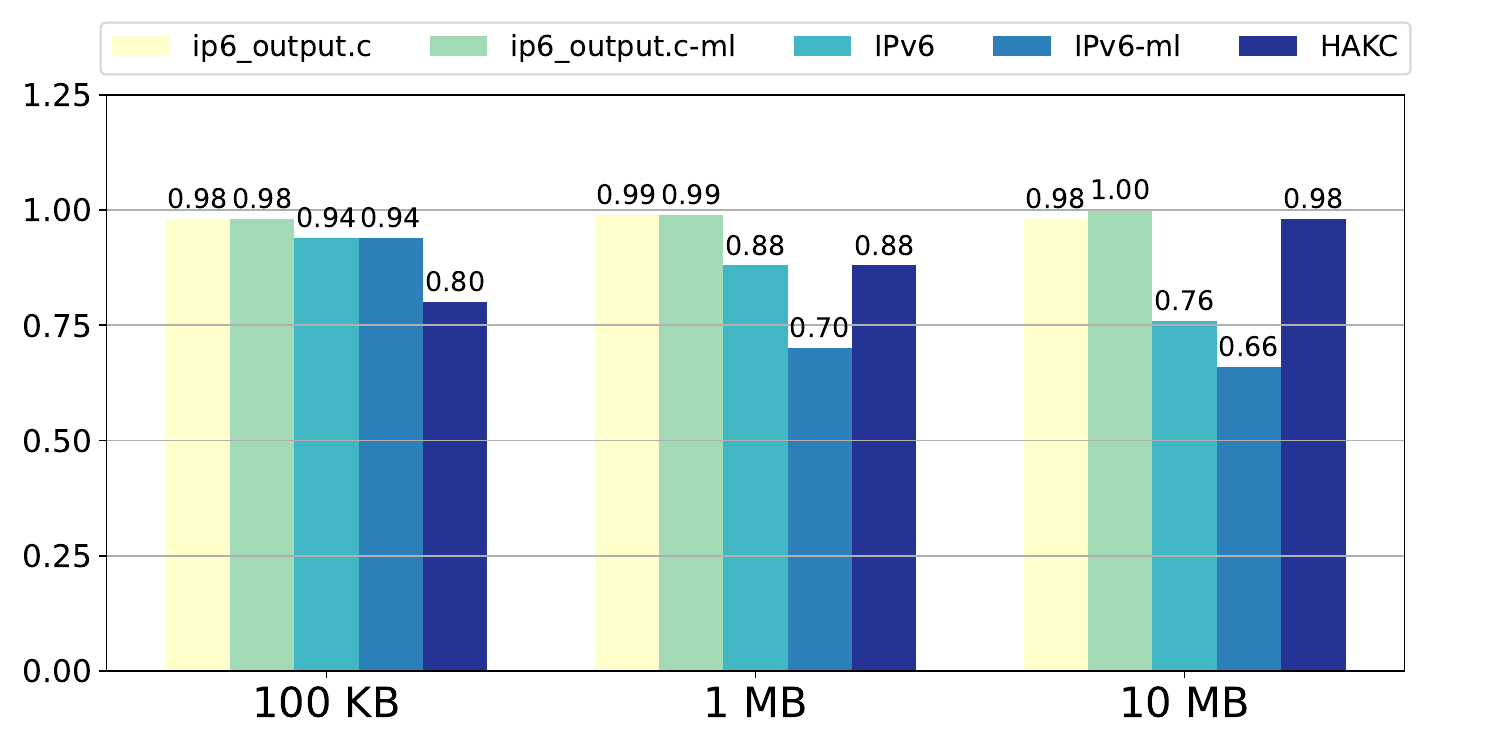}
         \vspace{-0.08in}
         \caption{Request/Sec}
         \label{fig:overhead-rps}
     \end{subfigure}
     \hfill
     \begin{subfigure}[b]{0.5\textwidth}
         \centering
         \includegraphics[width=\textwidth]{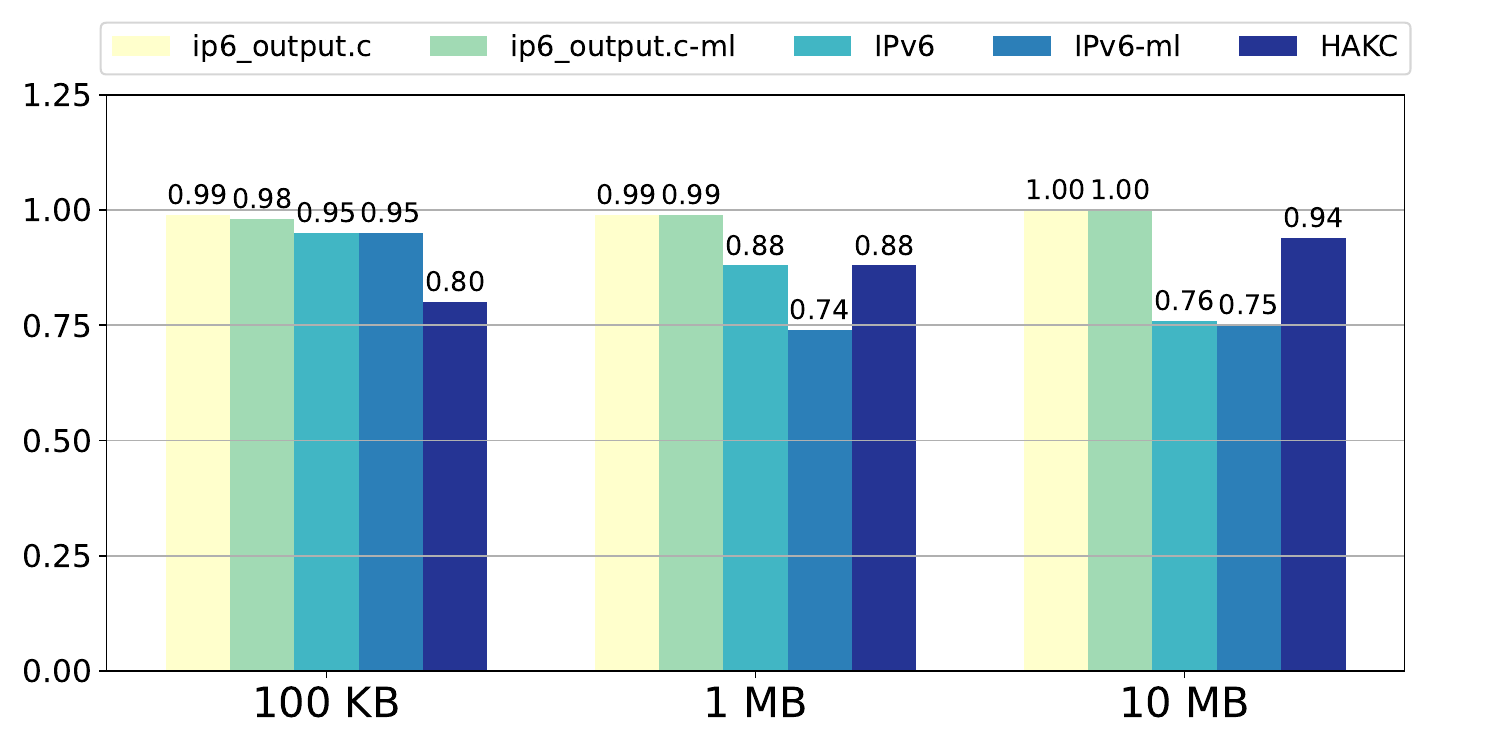}
         \vspace{-0.08in}
         \caption{Transfer Rate (Kbytes/Sec)}
         \label{fig:overhead-tr}
     \end{subfigure}
    \caption{Compartment-wise performance overhead, compared with the state-of-the-art HAKC (w.r.t., \textbf{RQ2}).}
    \label{fig:overhead}
\end{figure}

\vspace{0.05in}
\noindent \textbf{Results for RQ2.}
Figure~\ref{fig:overhead} presents the performance overhead imposed by \sys within the compartment. All measurement results are normalized to vanilla kernel.


From the figure, we can observe that the results for two metrics - Request/Sec and Transfer Rate - show similar trends.
More specifically, when the scale of the compartment is one single file, performance overhead is negligible for all file sizes (100KB, 1MB, and 10MB), even with ML audition.
However, if the scale is the entire IPv6 subsystem, performance decreases by 5\% for 100 KB files which is still tolerable, but sharply drops by nearly 20-35\% for 1MB and 10MB files.
At this scale, ML audition also exhibits impact on performance.
This change can be attributed to the file size.
The larger the file, the more frequent eBPF program executions become, and the more untracked objects in ML audition.



We compared \sys with HAKC.
Since the hardware feature used in HAKC, \ie, ARM MTE, is not publicly available yet, we can only use the experimental results reported in the paper for comparison.
From the figure, we can see that HAKC outperforms \sys when handling 10MB files.
This is because HAKC leverages the hardware features to eliminate checks when numerous memory accesses are in the same region.
However, the efficacy of such elimination diminishes with smaller file sizes.
That's why \sys shows advantages when handling 100KB files.
Beyond performance comparison, it is worth mentioning that HAKC lacks support for on-the-fly compartmentalization which is the main contribution of \sys.


%% file: tables/vulns.tex
\begin{table}[t]
\centering
\resizebox{\columnwidth}{!}{%
\begin{tabular}{|l|p{2.4cm}|l|}
\hline
\textbf{Title} & \textbf{Root Cause} & \textbf{Countermeasures} \\ \hline
CVE-2021-3715 & \begin{tabular}[c]{@{}l@{}}use-after-free in \\ \Code{sched/cls_route.c}\end{tabular} & \multirow{3}{*}{\begin{tabular}[c]{@{}l@{}}\sys eliminate \\ compartments to access\\  other kernel objects.\end{tabular}} \\ \cline{1-2}
CVE-2022-2588 & \begin{tabular}[c]{@{}l@{}}use-after-free in \\ \Code{sched/cls_route.c}\end{tabular} &  \\ \cline{1-2}
CVE-2022-27666 & \begin{tabular}[c]{@{}l@{}}heap overflow in \\ \Code{ipv6/esp6.c}\end{tabular} &  \\ \hline
CVE-2023-0394 & \begin{tabular}[c]{@{}l@{}}null pointer in \\ \Code{ipv6/raw.c}\end{tabular} & \multirow{2}{*}{\begin{tabular}[c]{@{}l@{}}\sys check the memory \\access target, avoid null\\ pointers/illegal addresses\end{tabular}} \\ \cline{1-2}
syz-76d0b8~\cite{syz-76d0b8049} & \begin{tabular}[c]{@{}l@{}}\#GP fault in \\ \Code{nft_tunnel.c}\end{tabular} &  \\ \hline
syz-e73923~\cite{syz-e7392359d} & \begin{tabular}[c]{@{}l@{}}stack overflow \\ in \Code{ipv6/sit.c}\end{tabular} & \multirow{2}{*}{\begin{tabular}[c]{@{}l@{}}compartment use its \\ own stack, return addresses \\ cannot be modified\end{tabular}} \\ \cline{1-2}
syz-490321~\cite{syz-4903218f7f} & \begin{tabular}[c]{@{}l@{}}integer overflow \\ in \Code{nfnetlink.c}\end{tabular} &  \\ \hline
syz-caed28~\cite{syz-caed28f292} & \begin{tabular}[c]{@{}l@{}}invalid free in \\ \Code{nf_tables_api.c}\end{tabular} & \begin{tabular}[c]{@{}l@{}}\sys check the kfree\\  arguments prevent \\ free other kernel objects\end{tabular} \\ \hline
\end{tabular}%
}
\vspace{0.05in}
\caption{Sampled vulnerabilities in IPv6, sched, and netfilter modules. Complete results are in ~\cite{alldata} due to the space limits.}
\label{tab:vulns}
\vspace{-0.3in}
\end{table}

%% file: tables/ml-compare-to-random.tex
\begin{table}[]
\centering
\resizebox{\columnwidth}{!}{%
\begin{tabular}{|ccccc|}
\hline
\multicolumn{1}{|c|}{\multirow{2}{*}{}} & \multicolumn{2}{c|}{\textbf{Type-Granularity}}                                       & \multicolumn{2}{c|}{\textbf{Compartment-Granularity}}                  \\ \cline{2-5} 
\multicolumn{1}{|c|}{}                  & \multicolumn{1}{c|}{\textbf{Accuracy}} & \multicolumn{1}{c|}{\textbf{Macro F1}} & \multicolumn{1}{c|}{\textbf{Accuracy}} & \textbf{Macro F1} \\ \hline
\multicolumn{5}{|c|}{\textbf{IPV6}}                                                                                                                                                    \\ \hline
\multicolumn{1}{|c|}{DT}                & \multicolumn{1}{c|}{96.88 ± 0.65}      & \multicolumn{1}{c|}{75.56 ± 1.84}      & \multicolumn{1}{c|}{99.99 ±0.02}       & 99.98 ± 0.03      \\ \hline
\multicolumn{1}{|c|}{RF}                & \multicolumn{1}{c|}{96.91 ± 0.63}      & \multicolumn{1}{c|}{78.81 ± 0.73}      & \multicolumn{1}{c|}{100 ± 0.01}        & 99.99 ± 0.01      \\ \hline
\multicolumn{1}{|c|}{NW}                & \multicolumn{1}{c|}{89.63 ± 1.29}      & \multicolumn{1}{c|}{38.76± 2.70}       & \multicolumn{1}{c|}{99.99 ± 0.01}      & 99.99 ±0.01       \\ \hline
\multicolumn{5}{|c|}{\textbf{net/sched}}                                                                                                                                                   \\ \hline
\multicolumn{1}{|c|}{DT}                & \multicolumn{1}{c|}{80.48 ± 0.76}      & \multicolumn{1}{c|}{71.04 ± 1.77}      & \multicolumn{1}{c|}{99.93 ± 0.14}      & 97.74 ± 4.22      \\ \hline
\multicolumn{1}{|c|}{RF}                & \multicolumn{1}{c|}{80.61 ± 0.69}      & \multicolumn{1}{c|}{76.28 ± 0.49}      & \multicolumn{1}{c|}{100 ± 0}           & 99.99 ± 0.01      \\ \hline
\multicolumn{1}{|c|}{NW}                & \multicolumn{1}{c|}{65.98 ± 6.91}      & \multicolumn{1}{c|}{39.18 ± 1.48}      & \multicolumn{1}{c|}{99.66±0.03}        & 89.47±1.20        \\ \hline
\multicolumn{5}{|c|}{\textbf{netfilter}}                                                                                                                                               \\ \hline
\multicolumn{1}{|c|}{DT}                & \multicolumn{1}{c|}{89.47 ± 0.23}      & \multicolumn{1}{c|}{78.17 ± 4.88}      & \multicolumn{1}{c|}{99.92 ± 0.07}      & 99.51 ± 0.46      \\ \hline
\multicolumn{1}{|c|}{RF}                & \multicolumn{1}{c|}{89.54 ± 0.15}      & \multicolumn{1}{c|}{81.87 ± 1.86}      & \multicolumn{1}{c|}{99.96 ± 0.05}      & 99.77 ± 0.29      \\ \hline
\multicolumn{1}{|c|}{NW}                & \multicolumn{1}{c|}{72.9 ± 2.23}       & \multicolumn{1}{c|}{37.98 ± 2.83}      & \multicolumn{1}{c|}{97.16 ±0.17}       & 74 ± 2.56         \\ \hline
\end{tabular}%
}
\vspace{0.05in}
\caption{Performance of the decision tree (DT) model customized for eBPF programs, the unrestricted random forest (RF), and the unrestricted neural network (NW) model.}
\label{ml tab:main}
\end{table}

%% file: tables/ml-parameters.tex

\begin{table}[]
\centering
\resizebox{\columnwidth}{!}{%
\renewcommand\arraystretch{1.2}
\begin{tabular}{|lcccc|}
\hline
\multicolumn{1}{|c|}{\multirow{2}{*}{\textbf{}}} & \multicolumn{2}{c|}{\textbf{Type-Granularity}}                                       & \multicolumn{2}{c|}{\textbf{Compartment-Granularity}}                  \\ \cline{2-5} 
\multicolumn{1}{|c|}{}                                & \multicolumn{1}{c|}{\textbf{Accuracy}} & \multicolumn{1}{c|}{\textbf{Macro F1}} & \multicolumn{1}{c|}{\textbf{Accuracy}} & \textbf{Macro F1} \\ \hline
\multicolumn{5}{|c|}{\textbf{Length of Feature Vector (Quad Word)}}                                                                                                                                      \\ \hline
\multicolumn{1}{|l|}{64}                              & \multicolumn{1}{c|}{89.15 ± 0.33}      & \multicolumn{1}{c|}{77.24 ± 4.21}      & \multicolumn{1}{c|}{99.91 ± 0.07}      & 99.47 ± 0.45      \\ \hline
\multicolumn{1}{|l|}{128}                             & \multicolumn{1}{c|}{89.18 ± 0.29}      & \multicolumn{1}{c|}{77.44 ± 4.33}      & \multicolumn{1}{c|}{99.85 ± 0.1}       & 99.46 ± 0.64      \\ \hline
\multicolumn{1}{|l|}{256}                             & \multicolumn{1}{c|}{89.26 ± 0.29}      & \multicolumn{1}{c|}{77.34 ± 5.06}      & \multicolumn{1}{c|}{99.92 ± 0.08}      & 99.51 ± 0.49      \\ \hline
\multicolumn{1}{|l|}{1024}                            & \multicolumn{1}{c|}{89.47 ± 0.23}      & \multicolumn{1}{c|}{78.17 ± 4.88}      & \multicolumn{1}{c|}{99.92 ± 0.07}      & 99.51 ± 0.46      \\ \hline
\multicolumn{5}{|c|}{\textbf{Maximum Depth}}                                                                                                                                                  \\ \hline
\multicolumn{1}{|l|}{3}                               & \multicolumn{1}{c|}{61.18 ± 2.45}      & \multicolumn{1}{c|}{1.72 ± 0.19}       & \multicolumn{1}{c|}{97.47 ± 0.4}       & 79.34 ± 3.03      \\ \hline
\multicolumn{1}{|l|}{7}                               & \multicolumn{1}{c|}{76.59 ± 2.38}      & \multicolumn{1}{c|}{8.48 ± 0.58}       & \multicolumn{1}{c|}{99.44 ± 0.21}      & 96.44 ± 1.32      \\ \hline
\multicolumn{1}{|l|}{10}                              & \multicolumn{1}{c|}{83.54 ± 2.19}      & \multicolumn{1}{c|}{21.06 ± 2.19}      & \multicolumn{1}{c|}{99.65 ± 0.14}      & 97.78 ± 0.86      \\ \hline
\multicolumn{1}{|l|}{14}                              & \multicolumn{1}{c|}{89.47 ± 0.23}      & \multicolumn{1}{c|}{78.17 ± 4.88}      & \multicolumn{1}{c|}{99.92 ± 0.07}      & 99.51 ± 0.46      \\ \hline
\end{tabular}%
}
\vspace{0.05in}
\caption{Performance of the decision tree model with varying parameters on netfilter subsystem. The results of other subsystems are consistent.}
\label{ml tab:depths}
\end{table}

%% file: tables/ml-dataset.tex
\begin{table}[t]
\centering
\resizebox{\columnwidth}{!}{%
\begin{tabular}{|l|l|r|}
\hline
\textbf{Data Source} & \textbf{Description} & \multicolumn{1}{l|}{\textbf{Entries}} \\ \hline
ipv6 & \begin{tabular}[c]{@{}l@{}}Objects for Common kernel services\\  + ipv6 protocal service\end{tabular} & 7,010,295 \\ \hline
net/sched & \begin{tabular}[c]{@{}l@{}}Objects for Common kernel services\\  +network packet scheduling and \\ Quality of Service (QoS)\end{tabular} & 6,190,592 \\ \hline
netfilter & \begin{tabular}[c]{@{}l@{}}Objects for Common kernel services\\  + network packet filtering and\\ Network Address Translation (NAT).\end{tabular} & 6,489,411 \\ \hline
\end{tabular}%
}
\vspace{0.05in}
\caption{Summary of training data for all ML models.}
\vspace{-0.1in}
\label{tab:ml-data}
\end{table}

\begin{table}[]
\centering
\resizebox{\columnwidth}{!}{%
\begin{tabular}{|l|l|p{6cm}|}
\hline
\textbf{Target} & \textbf{Type} & \textbf{Description} \\ \hline
\multirow{2}{*}{Object Type} & Content & The content of kernel objects collected by object profiler, divided into an array at 8-bytes granularity, each 8-bytes element represents a feature \\ \cline{2-3} 
 & Label & The type of the object \\ \hline
\multirow{2}{*}{Compartment} & Content & The content of kernel objects collected by object profiler, divided into an array at 8-bytes granularity, each 8-bytes element represents a feature \\ \cline{2-3} 
 & Label & Whether the object belongs to the compartment \\ \hline
\end{tabular}%
}
\vspace{0.05in}
\caption{Description of features and labels for all ML models.}
\vspace{-0.2in}
\label{tab:ml-features}
\end{table}

%% file: tables/overhead.tex
\begin{table}[]
\centering
\resizebox{\columnwidth}{!}{%
\renewcommand\arraystretch{1.2}
\begin{tabular}{lrrrrrr}
\hline
\multicolumn{1}{|c|}{}                                   & \multicolumn{2}{c|}{\textbf{\begin{tabular}[c]{@{}c@{}}ip6\_output.c \\ (1,470 LOC)\end{tabular}}} & \multicolumn{2}{c|}{\textbf{\begin{tabular}[c]{@{}c@{}}IPv6\\ (78,213 LOC)\end{tabular}}}    & \multicolumn{2}{c|}{\textbf{\begin{tabular}[c]{@{}c@{}}IPv6+\\sched+netfilter\\ (255,330 LOC)\end{tabular}}} \\ \cline{2-7} 
\multicolumn{1}{|c|}{\multirow{-2}{*}{\textbf{LMBench}}} & \multicolumn{1}{c|}{\textbf{ w  ML }}         & \multicolumn{1}{c|}{\textbf{w/o ML}}         & \multicolumn{1}{c|}{\textbf{ w  ML }}      & \multicolumn{1}{c|}{\textbf{w/o ML}}      & \multicolumn{1}{c|}{\textbf{ w  ML }}       & \multicolumn{1}{c|}{\textbf{w/o ML}}      \\ \hline
\multicolumn{1}{|l|}{Simple syscall}                     & \multicolumn{1}{r|}{-0.17\%}                    & \multicolumn{1}{r|}{-0.35\%}                    & \multicolumn{1}{r|}{-0.28\%}                 & \multicolumn{1}{r|}{-0.17\%}                 & \multicolumn{1}{r|}{-0.04\%}                  & \multicolumn{1}{r|}{-0.09\%}                 \\ \hline
\multicolumn{1}{|l|}{Simple read}                        & \multicolumn{1}{r|}{-0.15\%}                    & \multicolumn{1}{r|}{-0.01\%}                    & \multicolumn{1}{r|}{0.38\%}                  & \multicolumn{1}{r|}{-0.15\%}                 & \multicolumn{1}{r|}{0.02\%}                   & \multicolumn{1}{r|}{0.66\%}                  \\ \hline
\multicolumn{1}{|l|}{Simple write}                       & \multicolumn{1}{r|}{-0.23\%}                    & \multicolumn{1}{r|}{-0.12\%}                    & \multicolumn{1}{r|}{-0.07\%}                 & \multicolumn{1}{r|}{-0.23\%}                 & \multicolumn{1}{r|}{0.05\%}                   & \multicolumn{1}{r|}{-0.03\%}                 \\ \hline
\multicolumn{1}{|l|}{Simple stat}                        & \multicolumn{1}{r|}{0.44\%}                     & \multicolumn{1}{r|}{0.29\%}                     & \multicolumn{1}{r|}{1.63\%}                  & \multicolumn{1}{r|}{0.44\%}                  & \multicolumn{1}{r|}{0.47\%}                   & \multicolumn{1}{r|}{-0.91\%}                 \\ \hline
\multicolumn{1}{|l|}{Simple open/close}                  & \multicolumn{1}{r|}{1.79\%}                     & \multicolumn{1}{r|}{1.08\%}                     & \multicolumn{1}{r|}{1.58\%}                  & \multicolumn{1}{r|}{1.79\%}                  & \multicolumn{1}{r|}{3.45\%}                   & \multicolumn{1}{r|}{2.40\%}                  \\ \hline
\multicolumn{1}{|l|}{Select on tcp fd's}                 & \multicolumn{1}{r|}{-0.05\%}                    & \multicolumn{1}{r|}{-0.29\%}                    & \multicolumn{1}{r|}{0.32\%}                  & \multicolumn{1}{r|}{-0.05\%}                 & \multicolumn{1}{r|}{-0.14\%}                  & \multicolumn{1}{r|}{-0.32\%}                 \\ \hline
\multicolumn{1}{|l|}{Signal handler}                     & \multicolumn{1}{r|}{0.06\%}                     & \multicolumn{1}{r|}{-0.06\%}                    & \multicolumn{1}{r|}{-0.27\%}                 & \multicolumn{1}{r|}{0.06\%}                  & \multicolumn{1}{r|}{-0.06\%}                  & \multicolumn{1}{r|}{0.93\%}                  \\ \hline
\multicolumn{1}{|l|}{Protection fault}                   & \multicolumn{1}{r|}{1.17\%}                     & \multicolumn{1}{r|}{-0.05\%}                    & \multicolumn{1}{r|}{-0.86\%}                 & \multicolumn{1}{r|}{1.17\%}                  & \multicolumn{1}{r|}{1.19\%}                   & \multicolumn{1}{r|}{0.45\%}                  \\ \hline
\multicolumn{1}{|l|}{Pipe latency}                       & \multicolumn{1}{r|}{-0.64\%}                    & \multicolumn{1}{r|}{-1.26\%}                    & \multicolumn{1}{r|}{3.32\%}                  & \multicolumn{1}{r|}{-0.64\%}                 & \multicolumn{1}{r|}{-0.41\%}                  & \multicolumn{1}{r|}{-0.92\%}                 \\ \hline
\multicolumn{1}{|l|}{UDP latency}                        & \multicolumn{1}{r|}{-0.66\%}                    & \multicolumn{1}{r|}{0.65\%}                     & \multicolumn{1}{r|}{3.81\%}                  & \multicolumn{1}{r|}{-0.66\%}                 & \multicolumn{1}{r|}{3.65\%}                   & \multicolumn{1}{r|}{1.45\%}                  \\ \hline
\multicolumn{1}{|l|}{TCP latency}                        & \multicolumn{1}{r|}{0.37\%}                     & \multicolumn{1}{r|}{1.38\%}                     & \multicolumn{1}{r|}{0.44\%}                  & \multicolumn{1}{r|}{0.37\%}                  & \multicolumn{1}{r|}{2.14\%}                   & \multicolumn{1}{r|}{1.31\%}                  \\ \hline \hline
\multicolumn{1}{|c|}{\textbf{Phoronix}}                  & \multicolumn{1}{c|}{\textbf{w/ ML}}         & \multicolumn{1}{c|}{\textbf{w/o ML}}         & \multicolumn{1}{c|}{\textbf{ w  ML }}      & \multicolumn{1}{c|}{\textbf{w/o ML}}      & \multicolumn{1}{c|}{\textbf{ w  ML }}       & \multicolumn{1}{c|}{\textbf{w/o ML}}      \\ \hline
\multicolumn{1}{|l|}{perf-bench}                         & \multicolumn{1}{r|}{0.13\%}                     & \multicolumn{1}{r|}{-0.35\%}                    & \multicolumn{1}{r|}{0.53\%}                  & \multicolumn{1}{r|}{-0.55\%}                 & \multicolumn{1}{r|}{0.39\%}                   & \multicolumn{1}{r|}{-0.72\%}                 \\ \hline
\multicolumn{1}{|l|}{Kernel Compilation}                 & \multicolumn{1}{r|}{3.06\%}                     & \multicolumn{1}{r|}{3.30\%}                     & \multicolumn{1}{r|}{3.25\%}                  & \multicolumn{1}{r|}{2.70\%}                  & \multicolumn{1}{r|}{2.43\%}                   & \multicolumn{1}{r|}{3.74\%}                  \\ \hline
\multicolumn{1}{|l|}{XZ Compression}                     & \multicolumn{1}{r|}{-1.65\%}                    & \multicolumn{1}{r|}{-1.58\%}                    & \multicolumn{1}{r|}{-0.83\%}                 & \multicolumn{1}{r|}{-1.75\%}                 & \multicolumn{1}{r|}{-2.06\%}                  & \multicolumn{1}{r|}{-0.46\%}                 \\ \hline
\multicolumn{1}{|l|}{OpenSSL}                            & \multicolumn{1}{r|}{2.04\%}                     & \multicolumn{1}{r|}{2.06\%}                     & \multicolumn{1}{r|}{1.30\%}                  & \multicolumn{1}{r|}{1.37\%}                  & \multicolumn{1}{r|}{0.22\%}                   & \multicolumn{1}{r|}{0.86\%}                  \\ \hline
\multicolumn{1}{|l|}{SQLite Speedtest}                   & \multicolumn{1}{r|}{3.73\%}                     & \multicolumn{1}{r|}{1.32\%}                     & \multicolumn{1}{r|}{-2.90\%}                 & \multicolumn{1}{r|}{5.32\%}                  & \multicolumn{1}{r|}{-2.55\%}                  & \multicolumn{1}{r|}{-1.97\%}                 \\ \hline
\multicolumn{1}{|l|}{Nginx-ipv4}                         & \multicolumn{1}{r|}{1.81\%}                     & \multicolumn{1}{r|}{2.39\%}                     & \multicolumn{1}{r|}{2.20\%}                  & \multicolumn{1}{r|}{2.66\%}                  & \multicolumn{1}{r|}{3.02\%}                   & \multicolumn{1}{r|}{2.40\%}                  \\ \hline
\multicolumn{1}{|l|}{Apache-ipv4}                        & \multicolumn{1}{r|}{3.22\%}                     & \multicolumn{1}{r|}{3.76\%}                     & \multicolumn{1}{r|}{3.93\%}                  & \multicolumn{1}{r|}{3.69\%}                  & \multicolumn{1}{r|}{7.23\%}                   & \multicolumn{1}{r|}{5.89\%}                  \\ \hline
\multicolumn{1}{|l|}{Git}                                & \multicolumn{1}{r|}{2.49\%}                     & \multicolumn{1}{r|}{1.99\%}                     & \multicolumn{1}{r|}{2.67\%}                  & \multicolumn{1}{r|}{1.80\%}                  & \multicolumn{1}{r|}{1.45\%}                   & \multicolumn{1}{r|}{1.46\%}                  \\ \hline
\end{tabular}%
}
\vspace{0.05in}
\caption{Sampled results of \sys system-wide performance overhead, with vanilla kernel as the baseline (w.r.t., \textbf{RQ1}). Complete results can be found in \cite{alldata}.}
\label{tab:overhead}
\end{table}

%% file: sections/discussion.tex
\section{Discussion and Future Works}
\label{sec:discussion}

In this section, we will describe how we plan to extend the eBPF ecosystem to allow embedding more machine learning models. We will also discuss the decisions we made when designing  \sys and the reasons behind. 

\vspace{0.05in}
\noindent\textbf{Extending eBPF for More ML Models.} 
Recall that the current eBPF ecosystem has certain constraints, such as lacking support for floating-point calculations.
These constraints pose challenges to embed machine learning models into eBPF programs.
In this work, we bypassed these constraints and explored to apply the decision tree model for kernel tasks. 

Following this promising direction, we plan to extend eBPF's capability to embrace more powerful machine learning techniques. 
On one hand, to resolve the fundamental challenge concerning floating-point calculation, we draw inspiration from Dune~\cite{adam2012dune}, proposing to build a kernel-space sandbox specifically engineered to facilitate floating-point operations.
This sandbox will host the model, allowing for internal computation and result output.
On the other hand, to accelerate the prediction and inferring process, we plan to enrich types of BPF maps to store diverse data structures utilized in various machine learning models.


\vspace{0.05in}
\noindent \textbf{Shared Objects.} 
Data sharing between the compartment and the kernel is complicated.
Objects co-owned by the compartment and the kernel can be organized in structures like linked list or hierarchically managed, as revealed in prior works like LVD~\cite{narayananlvds20} and KSplit~\cite{huangksplit22}.
While these works provide mechanisms to maintain multiple copies of shared objects and synchronize the data when required, determining whether the stored data is malicious or not  remains an open challenge thus far.

\sys follows the Principle of Least Privilege. 
Therefore, the compartment is deprived of accesses to irrelevant data, but to keep the compartment's normal functionality, accessing to certain resources, including the shared objects, is retained.
Though there exists a potential risk of a malicious compartment exploiting the kernel via shared objects, this threat is substantially mitigated in \sys through scrutinizing over memory accesses and authorization of arguments and return values, as described in Section~\ref{sec:analyzer}.


\vspace{0.05in}
\noindent\textbf{Performance Optimization.}
\sys is the first work achieving compartmentalization on the fly.
Though it introduces minimal overhead in general, its performance can be further optimized without sacrificing security.
To name a few ad-hoc methods, checks within loops can be hoisted and redundant checks for individual objects can be further eliminated.
There are also academic works, such as Carat~\cite{suchy2020carat} and CaratCake~\cite{suchy2022caratcake}, provide more systematic solutions.

Furthermore, it isn't always necessary to compartmentalize the entire subsystem, oftentimes, selectively compartmentalizing security-critical files is sufficient.
Our statistics~\cite{alldata} reveals an uneven distribution of vulnerabilities across files in a kernel subsystem.
For example, in the net/sched module, 22 out of 57 discovered vulnerabilities are in a single file nf\_table\_api.c; in the IOuring module, 52 out of 85 discovered vulnerabilities are in 
io\_uring.c.
Compared to compartmentalizing an entire subsystem, the overhead of compartmentalizing a single file is negligible, as indicated in Section~\ref{sec:evaluation}.

Since performance optimization is not the primary focus of this work, we leave the implementation of aforementioned approaches as future works. 

%% file: sections/conclusion.tex
\section{Conclusion}
This paper presents \sys, a system that pioneers to embed machine learning models into the kernel space, leveraging the newest advancements in the eBPF ecosystem. 
\sys is the first work that achieves on-the-fly compartmentalization.
Our comprehensive evaluation validates that it effectively confines damage within the compartment, while imposing negligible overhead and showing excellent scalaibility system-wide.
Further, through comparative experiment, we demonstrate that decision tree is the most fitting machine learning model for \sys, due to is advantages in processing tabular data, its explainable nature, and its compliance with the eBPF ecosystem.
